# High-Performance Hybrid Silicon and Lithium Niobate Mach–Zehnder Modulators for 100 Gbit/s and Beyond

Mingbo He[1], Mengyue Xu[1], Yuxuan Ren[2], Jian Jian[1], Ziliang Ruan[2], Yongsheng Xu[2], Shengqian Gao[1], Shihao Sun[1], Xueqin Wen[2], Lidan Zhou[1], Lin Liu[1], Changjian Guo[2], Hui Chen[1], Siyuan Yu[1], Liu Liu[2,] and Xinlun Cai[1]

[1] State Key Laboratory of Optoelectronic Materials and Technologies and School of Physics and Engineering, Sun Yat-sen University, Guangzhou 510275, China.

[2] Centre for Optical and Electromagnetic Research, Guangdong Provincial Key Laboratory of Optical Information Materials and Technology, South China Academy of Advanced Optoelectronics, Sci. Bldg. No. 5, South China Normal University, Higher-Education Mega-Center, Guangzhou 510006, China.

Correspondence and requests for materials should be addressed to X. C. (caixlun5@mail.sysu.edu.cn) or to L.L. (email: liu.liu@coer-scnu.org).

Optical modulators are at the heart of optical communication links. Ideally, they should feature low insertion loss, low drive voltage, large modulation bandwidth, high linearity, compact footprint and low manufacturing cost. Unfortunately, these criteria have only been achieved on separate occasions. Based on a Silicon and Lithium Niobate hybrid integration platform, we demonstrate Mach–Zehnder modulators that simultaneously fulfill these criteria. The presented device exhibits an insertion loss of 2.5 dB, voltage-length product of 2.2 V·cm, high linearity, electro-optic bandwidth of at least 70 GHz and modulation rates up to 112 Gbit/s. The high-performance modulator is realized by seamless integration of high-contrast waveguide based on Lithium Niobate - the most mature modulator material - with compact, low-loss silicon circuits. The hybrid platform demonstrated here allows for the combination of "best-in-breed" active and passive components, opening up new avenues for enabling future high-speed, energy efficient and cost-effective optical communication networks.

Global data traffic has witnessed continuous growth over the past three decades due to the insatiable demand of modern society[1]. This rapid expansion is placing a serious challenge on the transceivers in all levels of optical networks, i.e., how to significantly increase data rates while reducing energy consumption and cost[2,3]. To address this challenge, silicon photonics on the silicon-on-insulator (SOI) platform has emerged as the leading technology due to the possibility of low-cost and high-volume production of photonic integrated circuits (PICs) in CMOS foundries[4-8].

Optical modulators are key components serving as the information encoding engines from the electrical domain to the optical domain[5]. Optical modulation in silicon mainly relies on free-carrier dispersion effect[9-14]. Unfortunately, free-carrier dispersion is intrinsically absorptive and nonlinear, which degrades the optical modulation amplitude (OMA) and may lead to signal distortions when using advanced modulation formats.

Tremendous efforts have been made towards realizing high-performance optical modulators in various material platforms[15-26]. Among them, Lithium Niobate (LN) remains a preferred material due to its excellent electro-optic (EO) modulation properties originating from Pockels effect[27-28]. LN modulators show unrivaled results for the generation of high-baud-rate multilevel signals and are still the best choice for ultra-long-haul links[29]. Conventional LN modulators are formed by low-index-contrast waveguides with weak optical confinement and the microwave electrodes must be placed far away from the optical waveguide to minimize absorption losses, which leads to an increased drive voltage. As a result, conventional LN modulators are bulky in size and low in modulation efficiency ($V_\pi$ L>10 V·cm). Recently, LN membranes on insulator (LNOI) has emerged as a promising platform to form waveguide devices with good confinement[29-40], and LNOI modulators with a low drive voltage and ultra-high EO bandwidth have been recently demonstrated[39,41-42].

An alternative approach, i.e., hybrid integration of LN membranes onto SOI PICs, has also attracted considerable interests[43-44]. The hybrid silicon/LN material system combines the scalability of silicon photonics with excellent modulation performance of LN. A few demonstrations of hybrid Si/LN optical modulators have been reported, and they all rely on a supermode waveguide structure consisting of an unpatterned LN membrane on top of a silicon waveguide. This structure is designed to support a distributed optical mode that resides in both the LN and the underlying silicon waveguide, i.e. only part of the modal power overlaps with the LN region, which compromises the modulation efficiency. In fact, the hybrid Si/LN optical modulators demonstrated so far show either low EO bandwidth or high operation voltage.

Here, we demonstrate hybrid Si/LN Mach-Zehnder modulators (MZMs) that employ two layers of hybrid integrated waveguides and vertical adiabatic couplers (VACs). The VACs transfer the optical power fully, rather than partially, between the silicon waveguide and LN membrane waveguide. This approach efficiently utilizes the LN membrane and overcomes the trade-off in the previous approaches. The proposed devices show a large EO bandwidth, high modulation efficiency, low on-chip insertion loss and high linearity. On-off keying (OOK) modulation up to 100 Gbit/s and four-level pulse amplitude modulation (PAM-4) up to 112 Gbit/s are successfully demonstrated.

## Results

**Design of Hybrid Silicon and LN MZMs.** The devices were fabricated based on benzocyclobuten (BCB) adhesive die-to-wafer bonding and LN dry etching techniques. Fig. 1(a) shows the schematic view of the present Si/LN hybrid MZM, consisting of two waveguide layers and VACs. The top waveguides, formed by dry-etching of an X-cut LN membrane, serves as phase modulators where EO interactions (Pockels effect) occur. The bottom SOI circuit supports all other passive functions, consisting of two 3 dB multimode interference (MMI) couplers that split and combine the optical power, and two grating couplers for off-chip coupling. The VACs, which were formed by silicon inverse tapers and superimposed LN waveguides, serve as interfaces to couple light up and down between the two layers. This hybrid integration architecture offers two distinct advantages. Firstly, since the task of routing light across the chip is placed on the underlying silicon waveguides, only simple, straight waveguides need to be fabricated in LN membrane. This allows for a more compact device footprint and greater flexibility in the LN waveguide design, compared with that of devices based on the pure LNOI platform. Secondly, the VACs, together with the dry-etched LN waveguide design, facilitate high overlap between the optical modes and active material, as well as good optical confinement in LN waveguide. This enables more efficient utilization of the LN active region, in contrast to other hybrid Si/LN hybrid devices with unpatterned LN membrane. The schematic of the cross-section of the hybrid waveguide is illustrated in Fig. 1(b).

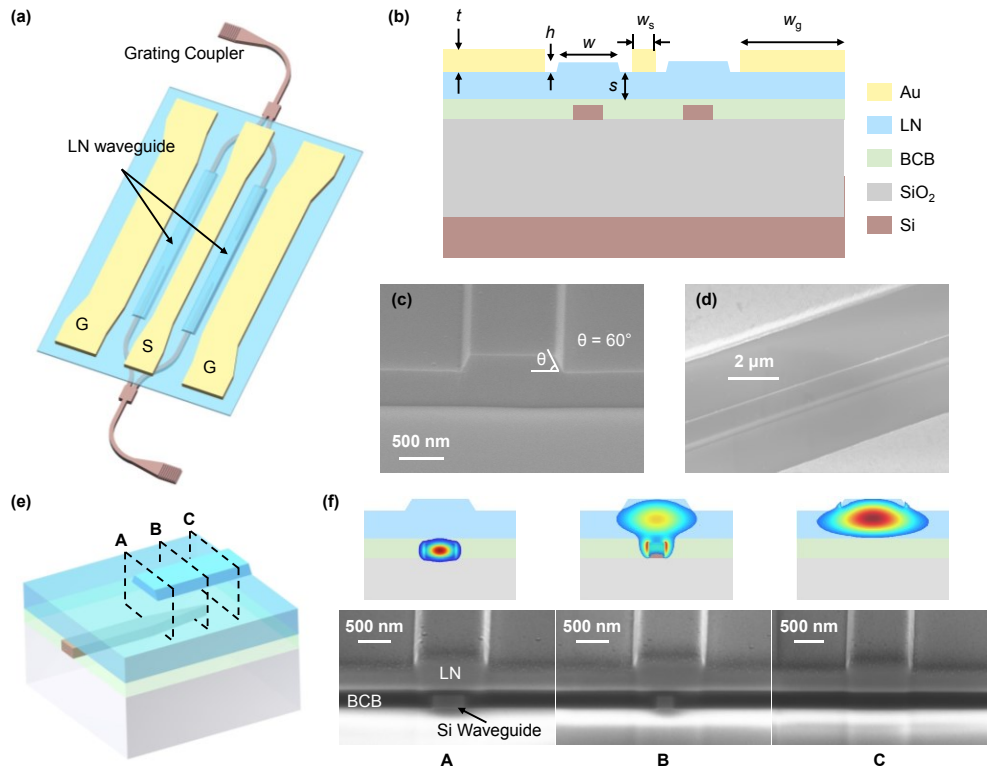

Figure 1 **Structure of the hybrid Si/LN MZM.** (a) Schematic of the structure of the whole circuit. (b) Schematic of the cross-section of the hybrid waveguide. (c) Scanning electron microscopy (SEM) image of the metal electrodes and the optical waveguide. (d) SEM image of the cross-section of the LN waveguide. (e) Schematic of the VAC. (f) SEM images of the cross-sections of the VAC at different positions, and calculated mode distributions associated with the cross-sections.

The LN waveguide is the most critical part of the present device and has to be optimized to achieve high modulation efficiency and low optical loss. The fabricated waveguides have a top width of $w = 1$ μm, a slab thickness of $s$=420 nm, a rib height of $h = 180$ nm and a sidewall angle of 60 degrees (Fig. 1(c)). The lithography and etching processes were optimized to yield smooth sidewalls and the LN waveguide features a measured propagation loss of 0.98 dB/cm (Supplementary Section III). The gap between the waveguides and electrodes was set to 2.75 μm. These parameters are designed to achieve good balance between the modulation efficiency and optical losses (including both metal absorption and sidewall scattering loss) (Supplementary Section I). The electrodes are configured in a ground-signal-ground (GSG) form, where the two LN waveguides lie in the two gaps between the ground and signal metals. To achieve large EO bandwidth, the electrodes are operated in a traveling wave manner and optimized for impedance matching, as well as velocity matching of the microwave and light signals (Supplementary Section II). The thickness of electrode was set to 600 nm, and the widths of signal and ground electrodes were designed to 19.5 μm and 30 μm, respectively (Fig. 1(d)).

The VAC is another important part of the device (Fig. 1(e)) and must be optimized for high efficiency. The width of the silicon waveguide tapers from 500 nm to 80 nm over a length of 150 μm, while the width of the LN waveguide remains constant at 1 μm. The thickness of the BCB is ~300 nm (i.e., the gap between the bottom of the LN waveguide and the top of the silicon waveguide is 80 nm), which is sufficiently robust with respect to the variations induced by the fabrication processes. The measured coupling efficiency of the VAC is > 97% (loss of ~0.19 dB per VAC) (Supplementary Section III). It should be noted that simulation indicates that the silicon taper tip width can be as large as 200 nm with negligible degradation in coupling efficiency, which means the present design is compatible with standard mass production processes (Supplementary Section III). Fig. 1(f) presents snap-shots of the optical intensity patterns for various cross-sections of the VAC, which illustrates the process of mode transfer.

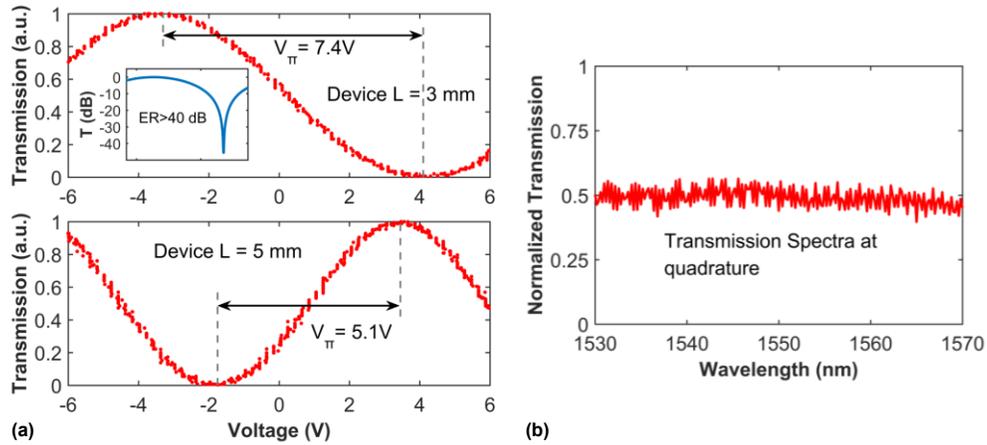

Figure 2 **Static EO Performance.** (a) Normalized optical transmission of the 3-mm and 5-mm device as a function of the applied voltage, showing $V_\pi$ of 4.5 V and 5.1 V, respectively. (b) Measured spectral response of the MZM biased at quadrature, indicating broadband operation of the device.

**Static EO Performance.** Fabricated MZM devices with arm lengths of 3 mm and 5 mm were measured in detail. The devices are driven in a push-pull configuration, so that applied voltage induces a positive phase shift in one arm and a negative phase shift in the other. Fig. 2(a) shows the half-wave voltage measurements for both devices with a 100-kHz triangular voltage sweep, and the half-wave voltages $V_\pi$ for the 3-mm and 5-mm devices are 7.4 V and 5.1 V, corresponding to voltage–length products of 2.2 V·cm and 2.5·V cm, respectively. The inset of Fig. 2 (a) shows the optical transmission on a logarithmic scale, indicating a measured extinction ratio of > 40 dB for the 3-mm device. Fig. 2(b) shows the measured spectral response of the MZM biased at quadrature, indicating broadband operation of the device in the whole C-band. The $V_\pi$ value of the present device can be further reduced by simply increasing the device length, as an LNOI modulator with $V_\pi$=1.4 V was recently achieved with a device length of 2 cm[39]. For the proposed Si/LN hybrid modulator, $V_\pi$ of about 1 V is expected with a similar length. This would enable driverless modulation from direct CMOS output without compromising the extinction ratio. In addition, a device with such a length can also fit in some common transceiver packages like QSFP (Quad Small Form-factor Pluggable) and can be adopted for future 400G applications[45].

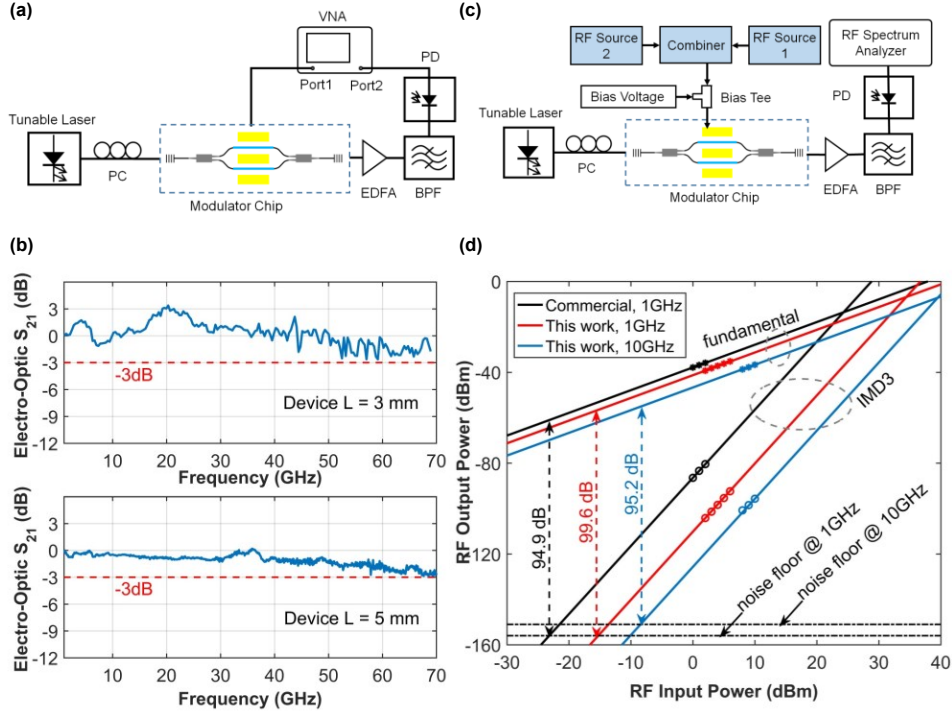

Figure 3 **EO Bandwidth and Linearity.** (a) Experimental setup for measuring the EO bandwidth. VNA, vector network analyzer; EDFA, erbium-doped fiber amplifier; BPF, band-pass filter; PD, photodetector; PC, polarization controller. (b) EO bandwidths ($S_{21}$ parameter) of MZMs with lengths of 3 mm and 5mm. The 3-dB bandwidths of both devices are beyond the measurement limit of the VNA (70 GHz). (c) Experimental setup for measuring the IMD3 SFDR. (d) RF output power of the fundamental and IMD3 components as a function of RF input power for our device at 1 GHz and 10 GHz, and for a commercially available LN MZM at 1 GHz. The noise floor is in 1 Hz bandwidth, limited by the RF spectral analyzer.

**EO Bandwidth and Linearity.** We then characterized the small signal EO bandwidth ($S_{21}$ parameter) of the fabricated devices using the setup shown in Fig. 3(a). The measured 3-dB EO bandwidths of both devices are > 70 GHz, which is beyond the measurement limits of our vector network analyzer (VNA). The measured EO bandwidth is much higher than that of pure silicon-based modulators. We believe that the EO bandwidth of the present device could be extended beyond 100 GHz by further optimizing the travelling-wave electrode (Supplementary Section II).

To characterize the linearity of the present device, we further examine the third order intermodulation (IMD3) spurious free dynamic range (SFDR) performance of the 3-mm device using the setup shown in Fig. 3 (c). To provide a comparative study, we also measured the IMD3 SFDR of a commercially available LN MZM (Fujitsu FTM7937EZ) with the same measurement system. Both devices are biased at quadrature and the optical power reaching the photodetector after pre-amplification is kept at 0 dBm. For our device, the measured IMD3 SFDR is 99.6 dB Hz$^{2/3}$ at 1 GHz, which is slightly better than that achieved by the commercial LN MZM (94.9 dB Hz$^{2/3}$). At 10 GHz, the measured SFDR decreases slightly to 95.2 dB Hz$^{2/3}$, mainly due to a higher noise floor of the measurement system. The SFDR value can be increased further by increasing the received optical power.

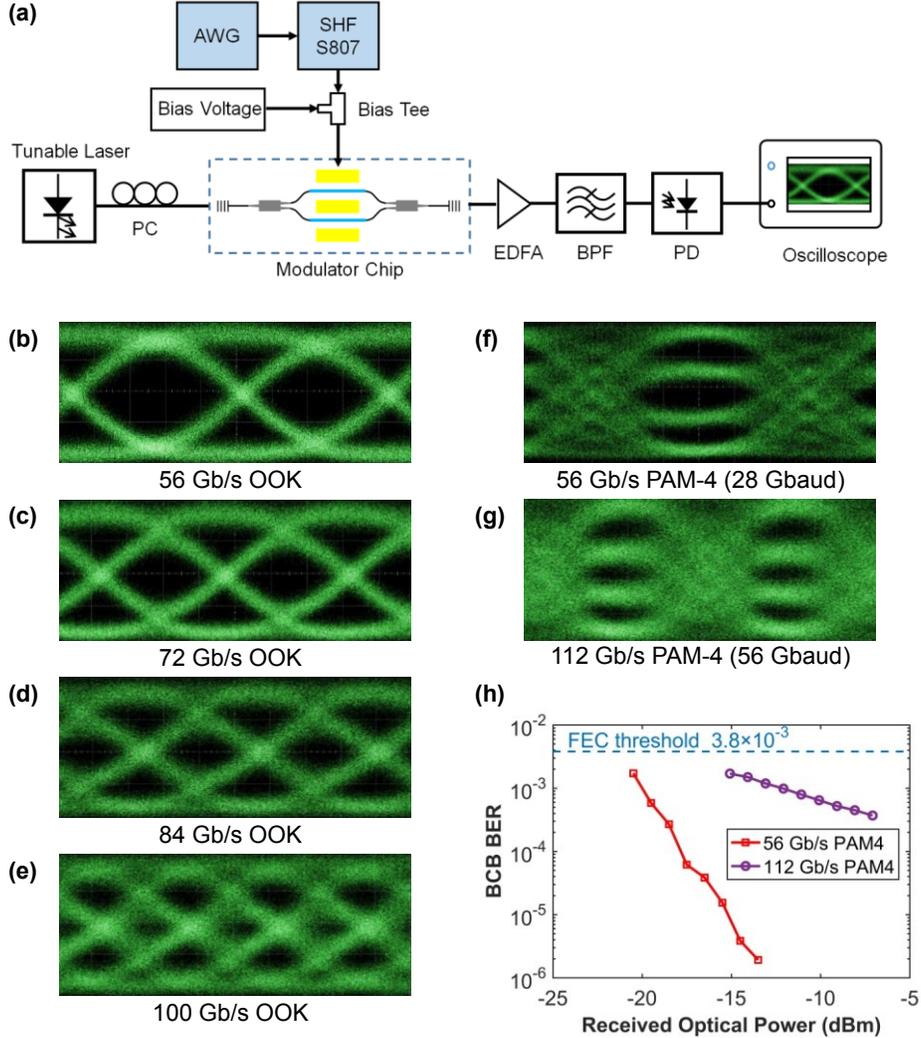

Figure 4 **Data Transmission Testing.** (a) Experimental setup for measuring the eye diagram. AWG, arbitrary waveform generator. (b-e) Optical eye diagrams for OOK signal at data rates of 56 Gb/s, 72 Gb/s, 84 Gb/s, and 100 Gb/s. The dynamic extinction ratios are 11.8 dB, 6.0 dB, 5.5 dB and 5.0 dB, respectively. (f-g) Measured PAM-4 modulation optical eye diagrams at 28 GBaud (50Gb/s) and 56 GBaud (112Gb/s). (h) Measured curves of BER versus the received optical power for 28 GBaud (56 Gb/s) and 56 GBaud (112 Gb/s) PAM-4 signal.

**Data Transmission Testing.** Next, we evaluated the performance of the 3-mm device for high-speed digital data transmission, as depicted in Fig. 4 (a). First, the OOK modulations were applied to the device. Fig. 4 (b-e) summarizes the optical eye diagrams at 56, 72, 84 and 100 Gb/s. The measured extinction ratios are 11.8, 6.0, 5.5 and 5.0 dB, respectively. It is noteworthy that at 100 Gb/s the whole measurement system is already limited by the bandwidth of the RF probe and cables. PAM-4 modulation experiments were also carried out with the same experimental setup. The optical PAM-4 eye-diagrams at 28 GBaud and 56 GBaud are shown in Fig. 4 (f-g). The back to back (B2B) bit-error rate (BER) curves at both PAM-4 data rates, shown in Fig. 4 (h), decrease linearly as the received optical power before pre-amplification increases. No error floor is observed in the measured power range, and the error rates are also well below the hard-FEC limit of $3.8 \times 10^{-3}$.

## Discussion

As presented above, the hybrid Si/LN MZMs demonstrated here can achieve excellent optical modulation characteristics, while maintaining key advantages of SOH photonic circuits. LN active waveguides can be bonded and fabricated with lithographic precision and alignment accuracy in a back-end process after the SOI fabrication. This manufacturing procedure is highly scalable and fully CMOS-compatible. Therefore, our approach potentially provides a new generation of compact, high performance optical modulators for telecommunications and data-interconnects.

It should be noted that approaches of combining silicon with other materials exhibiting Pockels effects, such as organic materials and barium titanate (BTO), have also been reported. In Tab. 1, we compare the performance of the present device with state-of-the-art, including silicon-organic-hybrid (SOH) MZMs, SOH plasmonic MZMs, and BTO/Si plasmonic MZMs. Here, we focus on MZMs with either large EO bandwidth or high operation speed, and results of pure silicon modulators operating in carrier depletion mode are also included as a benchmark. As shown in Tab. 1, the present device in this work is the only one, on which low $V_\pi$ and low insertion loss are achieved simultaneously. In particular, the insertion loss of the present device is much lower than that of all others, and to the best of our knowledge, this is the lowest insertion loss ever achieved in optical modulators operating above 40Gbit/s on silicon. The demonstrated $V_\pi$ of the present device is also promising, which is only surpassed by the SOH modulator with a much higher insertion loss. As discussed above, the $V_\pi$ can be further engineered to be about 1 V while keeping an insertion loss of less than 4 dB. A low $V_\pi$ is critical for a travelling-wave type of modulator, since the energy consumption per bit is proportional to the square of the driving voltage. For the data transmission experiments in Fig.4, the energy consumption of the present device of 3 mm length is estimated about 700 fJ/bit (see Supplementary Section II). By increasing the device length, it is possible to push the energy consumption down to about 40 fJ/bit while still keeping a bandwidth > 40 GHz. It has been recently proved that tens of aJ/bit energy consumption can be achieved for LN thin film modulators although at a compromised BER[39]. We believe that a similar performance can be expected for the present Si/LN hybrid device, since the cross-section structures of the phase shifter section are very similar for the two cases. Another approach for decreasing the energy consumption is to use a resonant device with lumped electrodes instead of a travelling-wave design[32,35], but the working wavelength band, in this case, is limited.

The underlying silicon waveguides not only route light with low losses across the chip, but also allow integration of the present modulator with the complete suite of silicon photonic components. This makes the proposed Si/LN platform highly desirable for new emerging applications. For instance, future integrated microwave photonic (MWP) systems would require on-chip linear modulators featuring high-fidelity electronic-to-optical conversion. The present device is ideal for this scenario and can be envisaged to be co-integrated with passive SOI structures, such as Bragg gratings or micro-resonators, to form sophisticated integrated MWP circuits. Another possible application area could be linear optical quantum computing (LOQC)[50]. LOQC relies on universal quantum gates which could be implemented by adding auxiliary photons and by using rapid, time-of-flight feedforward. In practice, a feedforward step requires large-scale, low loss, rapidly reconfigurable (in GHz range) optical circuits capable of low-noise pure phase modulation – requirements that are attainable on our platform. The present platform is also amenable to further integration with lasers and detectors, as well as high-speed electronic drivers based on CMOS technology, offering fully integrated solutions for future terabit-per-second datacom and interconnect applications. The fabrication process can also be adapted for materials such as $Si_3N_4$, which enables ultra-low loss PICs and potential platform for exploring nonlinear optical processes.

Table 1 Comparison of several performance metrics for hybrid silicon MZM

|  | Loss | $V_\pi$ | EO Bandwidth | OOK Data Rate | The Length of Modulation Area |
|---|---|---|---|---|---|
| SOI[10] | 5.4 dB | 7 V | 58 GHz | 90 Gb/s | 2mm |
| SOH[20] | 11 dB | 22 V | >100 GHz | N.A. | 0.5mm |
| SOH[46] | >11 dB[*] | 0.9 V | 25 GHz | 100 Gb/s | 1.1mm |
| SOH Plasmonic[47] | 12 dB | >40 V[**] | >65 GHz | 40 Gb/s | 29μm |
| Si/BTO Plasmonic[48] | 30 dB[***] | 25 V | >100 GHz | 72 Gb/s | 10μm |
| Si/BTO[49] | 3.3dB[****] | 20V | 800MHz | 300Mb/s | 0.75mm |
| This Work | 2.5 dB | 5.1 V | >70 GHz | 100 Gb/s | 5mm |

[*] The value is calculated from 20 dB fiber-to-fiber loss and 9 dB off-chip coupling loss.

[**] The value is calculated from the reported $V_\pi L$ of 1.3 Vmm and length of 29 μm.

[***] The value is calculated from propagation loss of 2 dB/μm with 10 μm length, and 10 dB loss for two photonic-plasmonic converters.

[****] The value is calculated from propagation loss of 44 dB/cm with 0.75 mm length.

**Methods**

**Fabrication:** A standard SOI processing including e-beam lithography (EBL) and dry etching processes was used to fabricate the grating coupler, 3dB MMIs, and silicon inverse tapers. An LNOI sample with silicon substrate was flip-bonded to the SOI wafer using BCB adhesive bonding process. The substrate of the LNOI was removed by mechanical grinding and dry etching. Afterwards, Hydrogen silsesquioxane (HSQ, FOX-16 by Dow Corning) was spin-coated on the LN thin film for EBL. The waveguide patterns were transferred into LN with optimized argon plasma etching in inductively coupled plasma (ICP) etching system. The physical etching by argon plasma results in a sidewall angle of ~60°. Finally, the travelling wave electrodes were patterned through a liftoff process.

**Dynamic measurement:** The setup for EO bandwidth, linearity, and data transmission measurements are shown in Fig. 3 and 4, respectively. The electrical signal is fed to the device electrode through a 67GHz-bandwidth RF probe (GGB 67A). A second RF probe (not shown) is also attached to the end of the travelling-wave electrode, and a 50 Ω termination is applied to the second probe for impedance matching. For the optical signal, light from a tunable laser is coupled into and collected from the SOI grating couplers using single mode fibers. A polarization controller is used to ensure TE input polarization. The received optical signal is pre-amplified and filtered through an erbium-doped fiber amplifier (EDFA) and a band-pass filter (BPF) before detection. A VNA is used for bandwidth measurement, and the frequency response of the PD (XPDV4120R) is deducted from the measured S21 response. For linearity measurement, the input electrical signal consists of a DC bias and two RF tones with frequencies separated by 10MHz and centered at the desired frequency. An RF spectrum analyzer is used to analyze frequency components of the received electrical signal. For data transmission measurement, an arbitrary waveform generator (AWG, Micram) with 100Gs/s sampling rate is used to generate the electrical signals. A 50 GHz broadband amplifier (SHF 807) with output saturation $V_{pp}$ of 4 V is used to amplifier the driving signal to the modulator together with a DC bias. For OOK eye diagram measurements, the modulator is biased at a voltage a little lower than the quadrature point to achieve better ERs. For PAM-4 eye diagram measurements, the modulator is biased at the quadrature point and the levels of the driving signals are deducted from an inverse trigonometric function in order to compensate the typical sinusoid response of the MZI. Digital low-pass filters are also applied to limit the signal bandwidth. For the bit-error rate measurements, the detected signal is sampled by a real-time oscilloscope (Tektronics DPO73304), and analyzed using an off-line DSP, including resampling, equalization, and symbol decision (see Supplementary Section IV).

### Author contributions

X.C. developed the idea. X.C. and L.L. conceived device design. M.H., J.J. carried out the LN fabrication. M.H., S.G., H.C., L.Z., L.L., and S.S. carried out the silicon fabrication. M.H. and Y.R. carried out the bonding process. M.X., Z.R., Y.X., X.W., C.G., carried out the measurement. L.L. and X.C. carried out the data analysis. All authors contributed to the writing. X.C. finalized the paper. S.Y., L.L., and X.C. supervised the project.

**Competing Interests** The authors have declared that no competing interests exist

# Supplementary Information

## High Performance Hybrid Silicon and Lithium Niobate Mach–Zehnder Modulators for 100 Gbit/s and Beyond

Mingbo He[1], Mengyue Xu[1], Yuxuan Ren[2], Jian Jian[1], Ziliang Ruan[2], Yongsheng Xu[2], Shengqian Gao[1], Shihao Sun[1], Xueqin Wen[2], Lidan Zhou[1], Lin Liu[1], Changjian Guo[2], Hui Chen[1], Siyuan Yu[1], Liu Liu[2] and Xinlun Cai[1]

[1] State Key Laboratory of Optoelectronic Materials and Technologies and School of Physics and Engineering, Sun Yat-sen University, Guangzhou 510275, China.

[2] Centre for Optical and Electromagnetic Research, Guangdong Provincial Key Laboratory of Optical Information Materials and Technology, South China Academy of Advanced Optoelectronics, Sci. Bldg. No. 5, South China Normal University, Higher-Education Mega-Center, Guangzhou 510006, China.

Correspondence and requests for materials should be addressed to X. C. (caixlun5@mail.sysu.edu.cn) or to L.L. (email: liu.liu@coer-scnu.org).

**I Design of LN waveguide**

It is essential that an LN phase modulator should have a low $V_\pi \cdot L$ and optical loss. The optical loss comes from two parts: the metal-induced absorption and sidewall scattering loss. In general, there is an inherent compromise between optical confinement and scattering loss from side-wall roughness. To maximize the modulation efficiency, a strong optical confinement, i.e., a deeply etched waveguide, is desired, because strong optical confinement facilitates a small electrode gap and low operation voltage. On the other hand, a waveguide that is too deeply etched will result in strong scattering from sidewall roughness, and the optical loss will also increase. In this section, we show how to optimize waveguide parameters for a low $V_\pi \cdot L$, and discuss the design considerations.

In order to calculate the $V_\pi \cdot L$, firstly, the electric field distribution ($E_z$) are computed in a commercial software (COMSOL Multiphysics). The distribution of the electric field of the present device ($w$=1μm and etch depth $h$=180nm) with an applied voltage of 1 V is shown in Fig. S1 (a). Then, we calculate the change of LN refractive index $\Delta n$, which is caused by the applied voltage, according to the formula:

$$\Delta n = \frac{E_z n_e^{3} r_{33}}{2} \tag{1}$$

where $E_z$ is the in-plane component of the electrical field in the LN waveguide, $n_e$ is the LN refractive index and $r_{33}$ is the electro-optic coefficient of LN.

Next, we calculate the distribution of the refractive index of the LN waveguide and use it in an optical mode solver (Lumerical Mode Solution) to find the fundamental TE mode at a wavelength of $\lambda_0$=1550nm. We define the effective refractive index change $\Delta n_{eff}$ as the difference between the effective refractive index of the fundamental TE mode with and without the applied voltage. $L = \frac{\lambda_0}{2\Delta n_{eff}}$ is the length at which the phase change reaches $\pi/2$ with an applied voltage of 1 V. Here, the phase change of $\pi/2$, instead of $\pi$, is used in the calculation, because the push-pull configuration in our device allows producing a $\pi$-phase shift between the two arms with only $\pi/2$ with opposite polarity in each arm. Therefore, $V_\pi \cdot L$ can be expressed as:

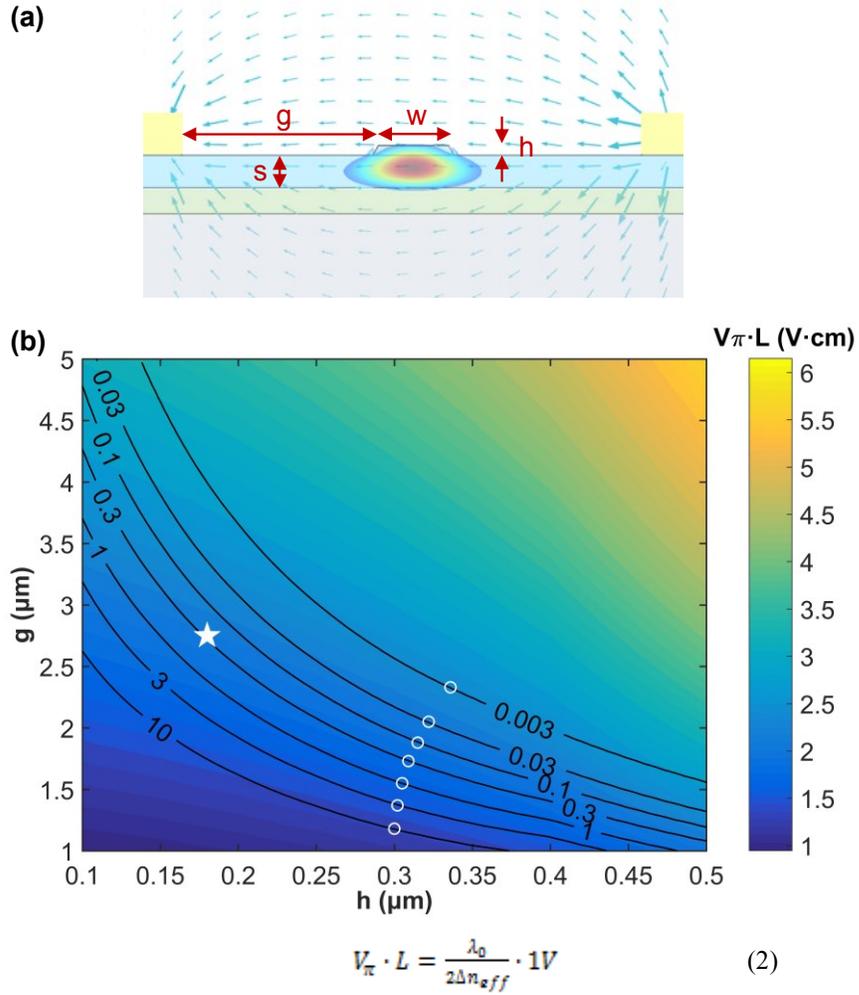

**(a)**

**(b)**

$$V_\pi \cdot L = \frac{\lambda_0}{2\Delta n_{eff}} \cdot 1V \qquad (2)$$

Figure S1. (a) A cross-sectional view of the simulated optical TE mode profile and the electric field (shown by arrows) between the two electrodes in this work. (b) Calculated $V_\pi \cdot L$ figures and mode losses induced by the metal absorption at difference $h$ and $g$. The color map is for $V_\pi \cdot L$ ($V \cdot cm$) and the black contour lines are for mode losses (dB/cm). The white circles indicate the positions for the lowest $V_\pi \cdot L$ on each contour line. The white star indicates the structural parameters used in the experiments of the main text.

The etching depth (the rib height of the LN waveguide) $h$ and the gap between the metal electrode and LN waveguide $g$ are crucial parameters for the optical performance of the present structure. Fig. S1 (b) shows the calculated $V_\pi \cdot L$ figures and mode losses induced by the metal absorption at difference $h$ and $g$. Intuitively, as the metal electrodes are put closer to the waveguide, the metal absorption would increase and $V_\pi \cdot L$ would decrease. On the other hand, as the etching depth becomes deeper, the absorption loss would decrease due to a better lateral confinement for the optical mode, while $V_\pi \cdot L$ would increase since the field residing in the LN material becomes less in this case. The above analyses are confirmed by the simulation data shown in Fig. S1(b). Furthermore, for each fixed metal absorption loss value, there exists an optimal set of structural parameters for the lowest $V_\pi \cdot L$. This conclusion is helpful for designing the present waveguide and electrode structure at a certain acceptable mode loss. It should be noted that only the metal absorption loss was analyzed here. Losses induced by other sources like sidewall roughness were not included. Actually, it is difficult to theoretical calculate the scattering loss from the sidewall roughness. Vast number of devices would be required in order to get a statistic map of the scattering losses related to structural parameters. This is under further investigation. Nevertheless, it can be believed that a shallow etched waveguide would present a lower scattering loss from the sidewall roughness. Therefore, in the experiments of the main text, we set the etching depth $h$ of 180 nm for the fabricated device. The gap $g$ is set 2.75μm. In this case, the calculated $V_\pi \cdot L = 2.46 \text{ V} \cdot \text{cm}$ and the metal induced absorption loss is below 0.3 dB/cm. It is not the lowest $V_\pi \cdot L$ value at the same loss level (the lowest $V_\pi \cdot L$ is obtained around $h$=300nm), but not too far off.

**II Design of Travelling Wave Electrodes and Energy Efficiency**

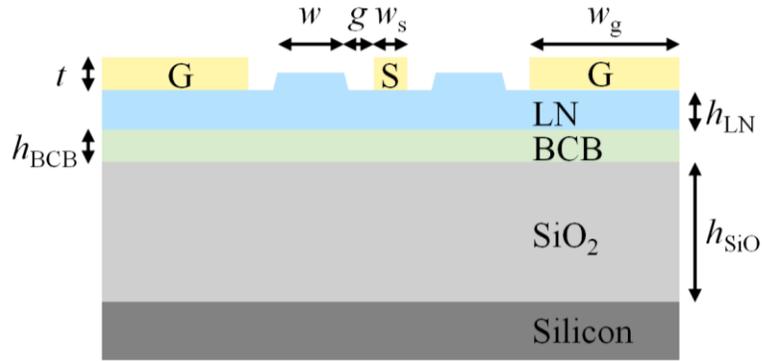

Figure S2. Schematic structure of the coplanar line waveguide in the arms of the present device.

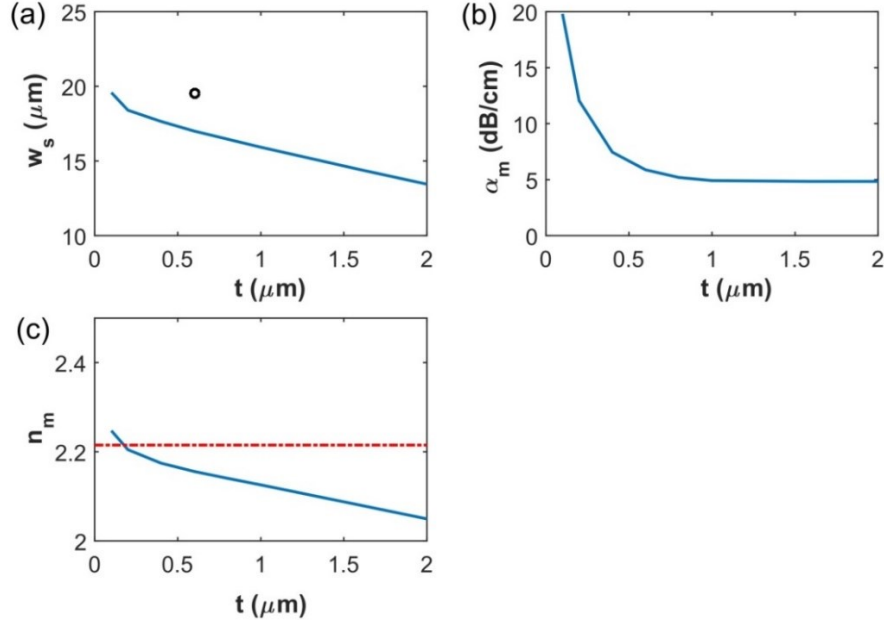

Figure S3. (a) Signal electrode width $w_s$, (b) RF attenuation $\alpha_m$, and (c) RF effective mode index $n_m$ of the coplanar waveguide at different electrode thicknesses $t$, while the real part of the characteristic impedance $Z_0$ is kept at 50Ω. The structural parameters adopted in the experiments discussed in the main text is marked in (a) as a circle. The group index of the optical mode is marked as a red dash-dotted line in (c).

The modulation performance of the present LN/Si modulator is analyzed in this section. It is known that the performance of a travelling wave modulator is related to the impedance of RF waveguide, the losses of the RF signal, and the propagation speed of optical and RF signals. Fig. S2 shows the cross-sectional structure at the arms of the MZM. For the RF signal, it is basically a coplanar line waveguide. Adopting the optical waveguide design from Sec. I in this document, some of the structural parameters are defined as $h_{SiO}$=3 μm, $h_{BCB}$=300 nm, $h_{LN}$=420 nm, and $w$+2$g$=6.5μm. We analyze the characteristics of the present coplanar line structure for different metal thicknesses $t$. For each metal thickness $t$, the width $w_s$ of the signal metal is first adjusted to ensure the characteristic impedance $Z_0$ of the waveguide to be 50 Ω at a frequency of 50 GHz. This relation is shown in Fig. S3(a). As expected, a thick metal would require a narrow signal metal to achieve impedance matching. Then, for each combination of $t$ and $w_s$, the RF effective mode index $n_m$ and RF attenuation $\alpha_m$ of the waveguide are further analyzed as shown in Fig. S3(b) & 3(c). For the optical mode, the group index $n_o$ of the hybrid LN waveguide is about 2.21, which is also marked in Fig. S3(b). Thanks to the LN membrane structure and the hybrid integration process, the propagation speed of the optical and RF signals is well matched in the present device. The index mismatch is below 6% for all the structural parameters considered in Fig. S3. This is in contrast to the ordinary bulk LN modulator [S1]. For the RF attenuation, one can find that a thick metal favors a low loss RF waveguide. The decreasing of the RF attenuation saturates when the metal thickness $t$ is larger than 1μm, which is also the preferred region for the metal thickness in order to achieve the ultimate performance of the present modulator. The structures adopted in the experiments discussed in the main text is also marked in Fig. S3(a). Apparently, the adopted structural parameters are not the optimum concerning the impedance matching and RF attenuation. The overall modulation frequency response $m(\omega)$ of a travelling wave modulator can be expressed as [S2]:

$$m(\omega) = \left| \frac{2z_{in}}{z_{in}+50} \frac{(50+z_0)F_+ + (50-z_0)F_-}{(50+z_0)e^{\gamma_m L} + (50-z_0)e^{-\gamma_m L}} \right|^2,$$

where, $\omega$ is the angular frequency of the RF signal, $Z_{in} = Z_0 \frac{50 + Z_0 \tanh(\gamma_m L)}{Z_0 + 50 \tanh(\gamma_m L)}$ is the impedance seen at the entrance of the electrode, $L$=3mm is the length of the electrode, $\gamma_m = \alpha_m + j \frac{\omega}{c_0} n_m$ is the RF propagation constant, and

$F_\pm = \left(1 - e^{\pm \gamma_m L - j \frac{\omega}{c_0} n_o L}\right) / \left(\pm \gamma_m L - j \frac{\omega}{c_0} n_o L\right)$. Here, the impedances of the source and load are all assumed 50 Ω. The RF mode properties and the frequency responses of two different structures, one for the structure adopted in the experiments ($t$=0.6μm and $w_s$=19.5μm) and the other for a further optimized structure ($t$=1.1μm and $w_s$=15.5μm), are plotted in Fig. S4. For a short device length of 3 mm, the experimented structure shows a similar frequency response to that of the optimized structure. Both of them give a 3-dB bandwidth well beyond 100GHz. In order to further decreasing the driving voltage of the present modulator, one can further lengthen the device. According to a simulated $V_\pi \cdot L$ =2.46 V·cm, a $V_\pi$ of 1.9 V can be obtained for a device with 13-mm long MZI arms. At such a device length, the experimented structure would only present a 3-dB bandwidth of 26 GHz. On the other hand, the bandwidth of the optimized structure can reach > 40 GHz, which can support 56 Gbaud PAM-4 data transmission.

The energy consumption of a travelling wave modulator can be estimated as $(V_{pp}/2)^2/(B \times 50)$ [S3], where $V_{pp}$ is the peak-peak voltage of the driving signal and $B$ is the bit rate. 50 Ω load is assumed here. For the data transmission discussed in Fig. 4 of the main text, $V_{pp}$ of 4 V is obtained after the RF amplifier, which is about half of the $V_\pi$ for the measured device (3 mm long). We can then calculate an energy consumption of 714 fJ/bit at B=112 Gb/s. According to the simulation in Fig. S4, a 13mm long device can give a $V_\pi$ of 1.9 V and ensure a bandwidth > 40GHz.

Assuming the same driving strength ($V_{pp}$ to $V_\pi$ ratio) as the 3-mm long device, i.e., 950 mV $V_{pp}$, and the same data rate, the energy consumption of the present device can be optimized to 40 fJ/bit.

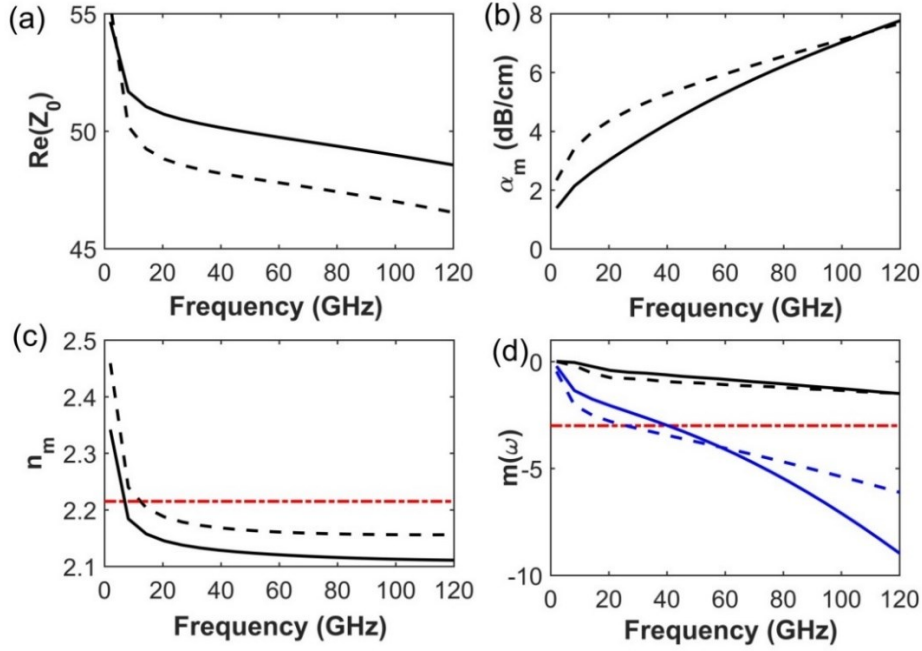

Figure S4. (a) Characteristic impedance $Z_0$, (b) RF attenuation $\alpha_m$, (c) RF effective mode index $n_m$, and (d) overall modulation response $m(\omega)$ of the present modulator with $t$=0.6 μm and $w_s$=19.5 μm (dashed lines) and $t$=1.1μm and $w_s$=15.5μm (solid lines). In (d), black lines are for 3mm long MZI arm length and blue lines are for 13mm. The red dash-dotted line marks -3dB position. In (c), the group index of the optical mode is marked as a red dash-dotted line.

### III Loss Measurement

To measure the propagation loss of the LN waveguide and the coupling efficiency of VAC, we designed and fabricated a series of meandering waveguide patterns containing silicon U-shaped bends, VACs and LN waveguides, as depicted in Fig. S5. The cross-section of LN waveguide is kept the same as the present modulator ( $w$= 1 μm and $h$= 180 nm).

Two grating couplers with 400-μm long linear tapers are used for off-chip coupling. Tapered waveguides are used to connect the grating couplers and single-mode silicon waveguides (500-nm wide). The loss of the grating couplers and the associated tapers are calibrated by the reference structures co-fabricated with the meandering waveguide patterns. Each meandering pattern is of the same number of U-shaped bends and VACs but with different lengths for LN waveguides for cut-back measurements. The number of the VACs is 62, and the lengths of LN waveguides are designed to be 0.62 cm, 1.86 cm, and 4.34 cm, respectively. The U-shaped bends are Euler bends with a minimum radius of 15 μm[S4], for which the bending loss is negligible. We use a cut-back approach to extract the propagation loss of the LN waveguide and the coupling efficiency of VACs. Fig. S6 shows the measured insertion losses of meandering waveguides with different lengths showing a linear propagation loss of 0.98 dB/cm. The coupling efficiency of VAC can be obtained from the intercept of the fitted curve and the Y-axis, and is estimated to be 8.02/62 = 0.13 dB (62 is the number of VACs in the pattern).

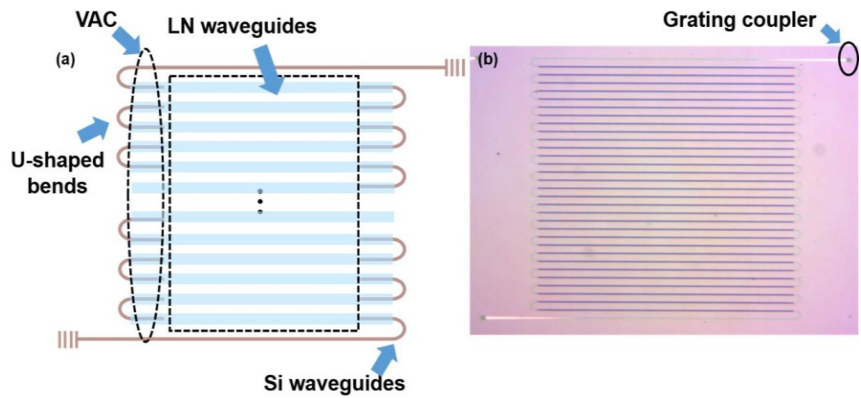

Figure S5. (a) Schematic diagram for measuring the insertion loss of LN waveguide and the coupling efficiency of VACs using cut-back method. (b) Microscope picture of the fabricated device

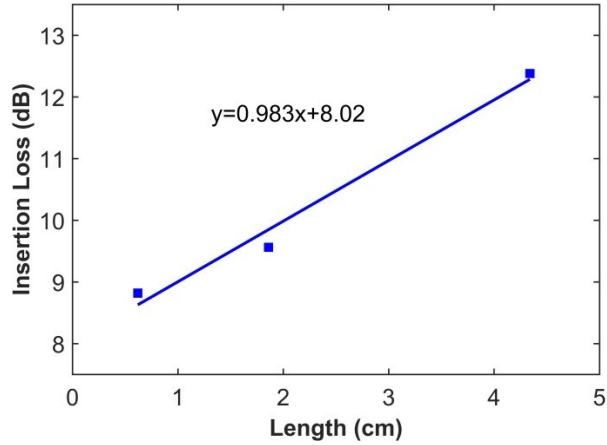

Figure S6. Measured insertion losses of LN waveguides with different lengths. The coupling efficiency of VACs can be calculated from the intercept of the fitted curve and the Y-axis.

We have also simulated the coupling efficiency of VACs against the width of the tip, and the result indicates that the width of the tip can be increased to be more than 200 nm with negligible degradation in coupling efficiency as shown in Fig. S7 (a). This means our design is compatible with standard CMOS tools. We have also simulated the influence of the alignment error on the coupling efficiency of VACs. As depicted in Fig. S7 (b), the alignment error can be as large as 300 nm, which is also compatible with standard lithography tools.

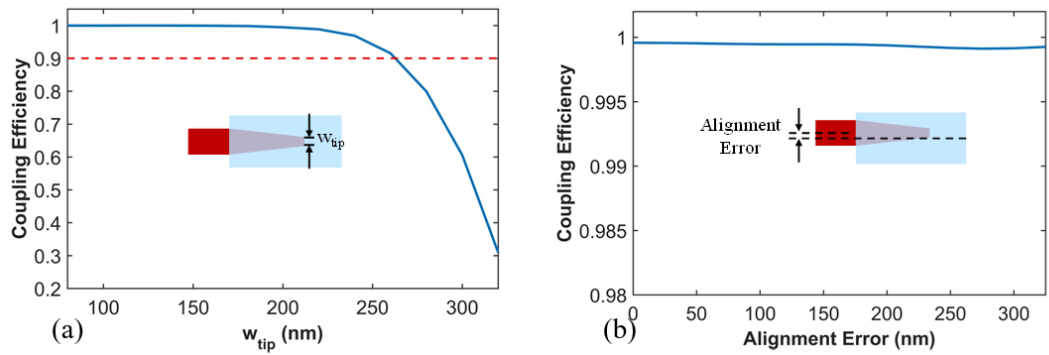

Figure S7. (a) Simulated coupling efficiency of VACs against the width of the silicon tip (b) Influence of the alignment error on the coupling efficiency of VACs

**IV DSP process for PAM-4 measurement.**

The DSP procedures used for offline processing of the received signals are depicted in Fig. S8. After sampled by a real-time oscilloscope (100-Gsa/s sampling rate, 33-GHz analog bandwidth), the received signal is firstly resampled to 2 samples/symbol (SPS). A feed-forward timing phase recovery algorithm [S5, S6] is then used to correct the timing error of the signal and generate an output signal with 1-SPS that sampled at the optimal sampling point. After timing phase recovery, a feed forward least mean square (LMS) equalizer with up to 31 taps is used for signal equalization. It can be seen from Fig. S9 that for 28-GBaud PAM-4 signals a 7-tap LMS is sufficient to get a satisfactory BER result, while for 56-GBaud signals the length of taps needed increases to 31, mainly due to the insufficient bandwidth of the real-time oscilloscope. Hard decision and error counting are then performed for BER evaluation. Note that all these DSP procedures are commonly used in commercial PAM-4 chipsets for short reach optical interconnect applications, with very low complexity [S7, S8].

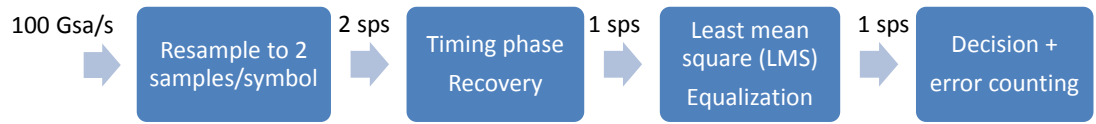

Figure S8. DSP block diagram for the received signals. sps: samples per symbol.

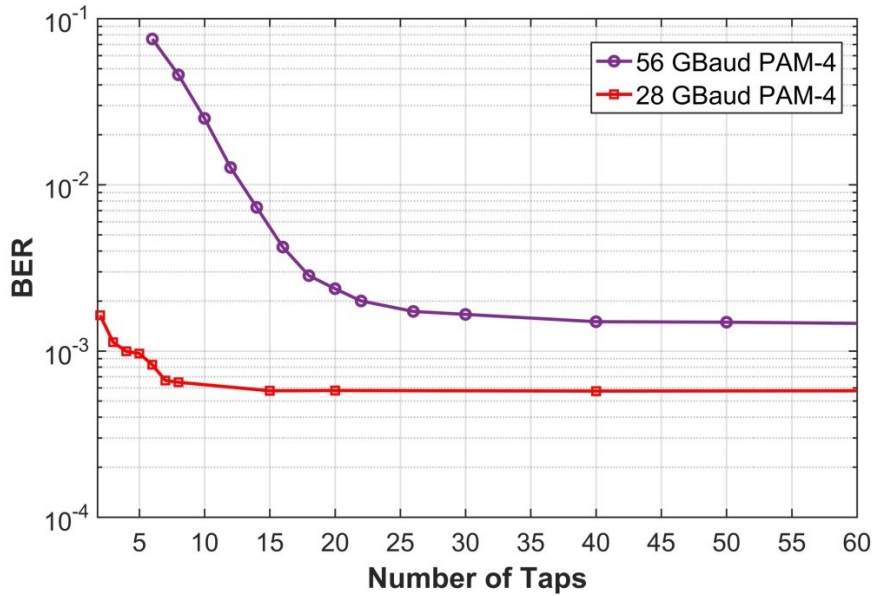

Figure S9. BER as a function of the number of taps used for LMS equalization.